\documentclass[aps,pre,showpacs,twocolumn]{revtex4}

\usepackage{epsfig,amsmath,amssymb,bm,epsf,graphics}
\usepackage{mathrsfs,txfonts}
\usepackage{float}

\begin{document}

\title{Molecular motion in cell membranes: analytic study of fence-hindered random
walks}
\author{V. M. Kenkre}
\affiliation{Consortium of the Americas for Interdisciplinary Science and
Department of Physics and Astronomy, University of New Mexico, Albuquerque,
New Mexico 87131, USA}

\author{L. Giuggioli}
\affiliation{Consortium of the Americas for Interdisciplinary Science and
Department of Physics and Astronomy, University of New Mexico, Albuquerque,
New Mexico 87131, USA}

\author{Z. Kalay}
\affiliation{Consortium of the Americas for Interdisciplinary Science and
Department of Physics and Astronomy, University of New Mexico, Albuquerque,
New Mexico 87131, USA}
\date{\today}

\begin{abstract}
A theoretical calculation is presented to describe the confined motion of transmembrane molecules in cell membranes. The study is analytic, based on Master equations for the probability of the molecules moving as random walkers, and leads to explicit usable solutions including expressions for the molecular mean square displacement and effective diffusion constants. One outcome is a detailed understanding of the dependence of the time variation of the mean square displacement on the initial placement of the molecule within the confined region. How to use the calculations is illustrated by extracting (confinement) compartment sizes from experimentally reported published observations from single particle tracking experiments on the diffusion of gold-tagged G-protein coupled $\mu$-opioid receptors in the normal rat kidney cell membrane, and by further comparing the analytical results to observations on the diffusion of phospholipids, also in normal rat kidney cells.
\end{abstract}

\maketitle

\section{Introduction}

Much activity has recently centered around the biophysics of cell membranes. This interest stems from the variety of processes in which the membrane plays a key role, among them cell shaping and movement \cite{mcmahongallop}, cell division \cite{celldiv}, signal transduction \cite{kraussbook}, and molecule trafficking \cite
{maxfieldtabas}. This latter process is of fundamental value for the
regulation of the localization, assembly, and aggregation of molecules
within, and in the vicinity of, the plasma membrane, functionalities that are
all linked to a healthy cell existence \cite{maxfieldtabas}. Understanding
the motion of membrane-associated molecules is, thus, of direct interest and great relevance.

Recent observations of lateral movements of molecules on the surface of the
cell \cite{edidin91,edidin94,sako98,tomishige,simson98,capps04,kusumireview} have led to the suggestion that the moving (transmembrane) molecules are confined within certain regions of
the cell membrane. This confinement is ascribed \cite{kusumireview} to collisions of membrane molecules protruding into the cytoplasm, the substance that mainly fills the cell interior, with the cytoskeleton \cite{howardbook}, a filament-like
structure present in the interior of the cell. The collisions are expected to reduce the movements of the molecules and
effectively confine their motion on the surface of the cell, as if the
plasma membrane were compartmentalized. Within each compartment, the
molecules are envisaged as moving freely, their motion being hampered as they traverse
adjacent compartments. As the actin filament that forms the compartment
boundary dissociates due to thermal fluctuations, a nearby transmolecule
with sufficient kinetic energy may overcome the barrier potential and hop
to the adjacent compartment. These arguments are the basis of the so-called
``membrane-skeleton fence model'' for temporary ``coralling'' of the transmembrane
molecules \cite{ritchieetal05}. Further slowing effects are also possible
due to ``transmembrane-protein pickets'' anchored on the membrane skeleton fence 
\cite{kusumireview}, in agreement with the latest research findings \cite
{engelman}, suggesting that the lipid bilayer plasma membrane is more mosaic
than previously thought \cite{singernicholson}. A recent publication by Morone et al. \cite{moroneetal} on imaging the cell surface by using  electron tomography techniques has provided further support for these models.

Observations \cite{kusumireview,ritchieetal05,muraseetal} of molecular
movements in the plasma membrane of various mammalian
cell types point to compartment dimensions between 30 and 240 nm and an
average hop rate between compartments ranging from 1 to 17 ms. Whereas diffusion constants of these molecules are known to be about 10 $\mu m^{2}s^{-1}$ in the absence of compartments, the above observations lead to the extraction of \emph{effective} diffusion constants 5 to 50 times smaller for the motion associated with hops between compartments \cite{kusumireview}. These experimental results have been analyzed \cite{kusumireview} in the past in terms of the model of Powles et al. \cite{powlesetal}, wherein the molecule is looked upon as a random walker moving in an infinite 1-D space with periodically arranged semipermeable barriers. The short-coming of the theory delineated in ref. \cite{powlesetal} is that the space-time dependence of the probability distribution of the random walker is obtained in terms of an unwieldy infinite series of terms which is difficult to handle. Furthermore the mean square displacement, which is the quantity of direct comparison to the experiment \cite{kusumireview}, is not derivable analytically from ref. \cite{powlesetal}.

Against this backdrop, which consists of experiments exhibiting non-standard (meaning non-free) diffusion of molecules on a cell membrane, and a theory which is not easy to validate, use, or manipulate, because of its involved expressions, we present a new analysis for hindered molecular diffusion on cell membranes. It is based on calculations carried out by one of the present authors, but not published, on the formally (but not physically) similar problem of Frenkel exciton transport \cite{kenkrebook} in deuterated molecular crystals. We will see that our theory results in easily usable expressions for the molecular mean square displacement and for the effective diffusion constant. We will show explicitly how to use our expressions by deducing reasonable values for the compartment size from experiments on cell membranes.

The paper is organized as follows: The model and its formal solution are in section \ref{model}, the continuum limit leading to the effective diffusion constant is in section \ref{continuum}, and explicit usable expressions for the time dependence of the molecular mean square displacement are in section \ref{expexp}. A comparison to the experiments of Kusumi and collaborators \cite{ritchieetal05} is given in section \ref{experiment} and concluding remarks constitute section \ref{conclusion}. A few calculational details may be found in the Appendix.

\section{The model and its formal solution in discrete space}
\label{model}
In constructing our model we follow Powles et al. \cite{powlesetal} in considering a random walker in an
infinite 1-D space but we study a discrete chain of sites with lattice constant $a$. Although, for the sake of simplicity, we display the
results only in 1-D as in earlier work \cite{powlesetal}, generalization to higher dimensions (e.g., 2-D) is straightforward at least in principle. The molecule, whose
probability of occupation of the $m$-th site of the chain at time $t$ is $%
P_{m}(t)$, hops with nearest neighbor rates $F$ within a compartment of $H+1$
average number of sites and with a lower rate $f$ from the end site of the
compartment to the first site of the adjacent compartment. For simplicity, $H$ is
taken to be even here, with the site $0$ at the center of one of the
compartments. The equation of motion is therefore 
\[
\frac{dP_{m}(t)}{dt}=F\left[ P_{m+1}(t)+P_{m-1}(t)-2P_{m}(t)\right] , 
\]
if $m$ is within a compartment, 
\[
\frac{dP_{r}(t)}{dt}=f\left[ P_{r+1}(t)-P_{r}(t)\right] +F\left[
P_{r-1}(t)-P_{r}(t)\right] , 
\]
for the end site of one compartment, and 
\[
\frac{dP_{r+1}(t)}{dt}=f\left[ P_{r}(t)-P_{r+1}(t)\right] +F\left[
P_{r+2}(t)-P_{r+1}(t)\right] , 
\]
for the first site of the next compartment. Here $r$ and $r+1$ mark the site
location of, respectively, the left and right ends of each compartment
boundary. The index $r$ equals $H/2+(H+1)l$ with $l$ an integer
running from $-\infty $ to $+\infty .$ 

With $\Delta =F-f$ , the three equations can be combined: 
\begin{eqnarray}
\frac{dP_{m}(t)}{dt}=F\left[ P_{m+1}(t)+P_{m-1}(t)-2P_{m}(t)\right] \nonumber \\ -\Delta \sum_{r}^{\prime}\left[ P_{r+1}(t)-P_{r}(t)\right] \left( \delta _{m,r}-\delta
_{m,r+1}\right)  \label{starting}
\end{eqnarray}
where the primed summation is over the left ends $r$ of all barriers
and the $\delta $'s are Kronecker deltas. Equation (\ref{starting}) is
formally identical to an equation studied in the past (see ref. \cite{kenkrebook} and references therein) to
describe exciton processes in a molecular crystal. Using tildes to denote Laplace
transforms and $\epsilon $ to denote the Laplace variable, it is possible to
solve  Eq. (\ref{starting}) as (see Appendix )
\begin{eqnarray}
\widetilde{P}_{m}(\epsilon )=\widetilde{\eta }_{m}(\epsilon )-\left(\frac{\Delta}{1+\Delta \widetilde{\mu }(\epsilon )}\right) \sum_{r}^{\prime}%
\left[\widetilde{\eta }_{r+1}(\epsilon )-\widetilde{\eta }_{r}(\epsilon )\right]\nonumber \\ \times \left[ \widetilde{\Psi }_{m-r}(\epsilon
)-\widetilde{\Psi }_{m-r-1}(\epsilon )\right] ,  \label{explicit}
\end{eqnarray}
where $\Psi _{mn}\left( t\right) =\Psi _{m-n}\left( t\right) $ is the
probability propagator for the system without barriers ($\Delta =0$), i.e.
the probability that the molecule is at site $m$ at time $t$ if it was at
site $n$ at time $0$ in the barrier-less system. The dependence of the
propagator on merely the difference in the site locations stems from
translational invariance. By $\eta _{m}\left( t\right) $ is meant the
solution $\sum_{n}\Psi _{m-n}\left( t\right) P_{n}\left( 0\right) $ of the
homogeneous (barrier-less) problem for the given initial probability
distribution $P_{n}\left( 0\right) $.

Information about the location of the barriers is in the crucial function $\widetilde{\mu }(\epsilon)$, which, (see Appendix), for periodic placement of barriers, is given by
\begin{equation}
\widetilde{\mu }(\epsilon
)=\frac{1}{F}\left[ \frac{\tanh \left( \xi /2\right) }{\tanh \left( \xi\left( H+1\right) /2\right) }-1\right] ,
\label{mueq}
\end{equation}
where $\cosh \xi=1+\epsilon/2F$. This closed form for $\widetilde{\mu }(\epsilon)$ is a consequence of the fact that the propagators for the barrier-less chain with nearest-neighbor rates, i.e., the system obeying Eq. (\ref{starting}) with $\Delta=0,$ are products of modified Bessel functions and exponentials in the time domain, specifically $\Psi_l(t)=I_l(2Ft)e^{-2Ft}$, the consequent Laplace transforms being $$\widetilde{\Psi_l}(\epsilon)=\frac{e^{-\xi|l|}}{2F\sinh \xi}.$$

Since the $\eta$'s are easily obtainable (being solutions of the homogeneous equation, as explained above), in light of Eq. (\ref{mueq}), Eq. (\ref{explicit}) constitutes the full solution for the probabilities. A quantity of direct contact to experiments on the cell membrane is the mean square displacement. Because of the presence of barriers, one has to take care in defining this quantity, focusing in particular on its dependence of where the molecule starts in relation to the barriers. Translational invariance of the compartments allows one to focus on any one compartment without loss of generality. We consider thus, as the most general case for a fully site-localized initial condition, the particle starting at the $p^{th}$ site inside the central compartment so that $P_{n}(0)=\delta_{n,p}$, $p$ being any integer in the interval $[-H/2,H/2]$. In this case, $\widetilde{\eta }_{m}=\widetilde{\Psi}_{m-p}$. Then the mean square displacement for a localized initial condition, defined as the average of $(m-p)^2$, $$y_p(t)=\sum_m(m-p)^{2}P_m(t),$$ and labeled by the starting site $p,$ is given in the Laplace domain by:
\begin{align}
	\widetilde{y_p}(\epsilon )=\sum_{m}(m-p)^{2}\widetilde{\Psi}_{m-p}(\epsilon ) \nonumber \\ -\Delta \sum_{r}^{\prime}\widetilde{p}_{r}(\epsilon )\sum_{m}(m-p)^{2}\left[ \widetilde{
\Psi }_{m-r}(\epsilon )-\widetilde{\Psi }_{m-r-1}(\epsilon )\right] ,
\end{align}
and has obvious dependence on the starting site $p.$ Here $p_r,$ defined and evaluated in the Appendix, is simply the difference $P_{r+1}-P_r.$ Since $m^{2}=\left( m-r\right)
^{2}+2r\left( m-r\right) +r^{2},$ we can write, using  the fact that $\sum_{m}\widetilde{\Psi }_{m}=1/\epsilon $ and that symmetry
considerations dictate that the first moment of the propagators
vanishes, 
\[
\sum_{m}(m-p)^{2}\left[ \widetilde{\Psi }_{m-r}(\epsilon )-\widetilde{\Psi }%
_{m-r-1}(\epsilon )\right] =-\frac{2r-2p+1}{\epsilon },
\]
obtaining the mean square displacement as the sum of its counterpart for the
barrier-less case and a correction which is
proportional to the difference of the rates $\Delta =F-f:$%
\begin{gather}
\widetilde{y}_p( \epsilon)=\widetilde{y}_p( \epsilon)^{barrierless} \nonumber \\ -\frac{1}{\epsilon }\left(\frac{\Delta }{1+\Delta 
\widetilde{\mu }(\epsilon )}\right)\sum_{r}^{\prime}\left( 2r-2p+1\right) \left[ \widetilde{\Psi}_{r-p}(\epsilon )-\widetilde{\Psi}_{r+1-p}(\epsilon )\right] .
\label{firstpartition}
\end{gather}
By defining a function $g_p(t)$ with the Laplace transform 
\begin{equation}
\widetilde{g}_p\left( \epsilon \right) =\frac{\Delta }{F}\frac{\epsilon}{1+%
\Delta \widetilde{\mu }}\sum_{r}^{\prime}\left( r-p+\frac{1}{2}\right)  \left( 
\widetilde{\Psi }_{r-p}-\widetilde{\Psi }_{r+1-p}\right) ,  \label{g}
\end{equation}
and writing $\phi_p \left( t\right) =\delta \left( t\right) -g_p(t)$, we have
\[
y_p(t)=2F\int_{0}^{t}dt^{\prime
}\int_{0}^{t^{\prime }}\phi_p \left( t^{\prime \prime }\right) dt^{\prime
\prime }
\]
where $\phi_p(t)$ may be called the memory function \cite{kenkrebook}.

One of the advantages of expressing the mean square displacement in terms of the memory is that, for any starting site $p,$ the effective hopping rate $F_{eff}$ at large times can be easily calculated as $F_{eff}=F\int_{0}^{+\infty }dt\phi_p (t)$ and is given by 
\begin{equation}
F_{eff}=F\widetilde{\phi_p }\left( 0\right) =f\left[ \frac{H+1}{1+\left(
f/F\right) H}\right] .  \label{Feffective}
\end{equation}
That $\widetilde{g_p }(0),$ and consequently $\widetilde{\phi_p }(0)$, is independentt of $p$ will be clear below. Equation (\ref{Feffective}) shows that if $f=F,$ there is no confinement
effect and $F_{eff}=F.$ If the intercompartmental motion is absent,
i.e., if $f=0,$ the molecule cannot escape and $F_{eff}=0.$ For this case the
mean square dispacement saturates to within the compartment. For small
intercompartmental escape, i.e., if $f<<F,$ specifically if we can
neglect $\left( f/F\right) H<<1,$ then $F_{eff}=f\left( H+1\right) .$ This
means that the molecule jumps large distances of compartment size at the
lower rate $f.$ We also see, the prediction of little interest to the cell
membrane problem but interesting in other contexts, that if on arrival at a
barrier the molecule is whisked away into the next compartment, specifically
if $f>>F/H,$ there is a small enhancement of the motion rate: $%
F_{eff}=F\left( 1+\frac{1}{H}\right) $which cannot exceed a factor of $3/2$.

In the Appendix  it is shown that the evaluation of Eq. (\ref{g}) gives
\begin{align}
\widetilde{g}_p(\epsilon)&=\frac{\Delta}{F}\frac{1}{P(\epsilon)+\left( \epsilon/2F+2f/F \right)Q(\epsilon)}\nonumber \\ &\times \bigg[  \left(H+1\right)\frac{P(\epsilon)+\left(\epsilon/2F\right) Q(\epsilon)}{P(\epsilon)+\left( 2+\epsilon/2F \right)Q(\epsilon)}\cosh(\xi p)  \nonumber \\ &-2p\sinh(\xi p)\tanh(\xi/2) \bigg]
\label{gepsfull}
\end{align}
where $P(\epsilon )=\cosh ( \xi H/2)$ and $Q(\epsilon )=\sinh ( \xi H/2) /\sinh \xi $. Note that $\widetilde{g}_p$ is symmetric in the sign of the initial site $p$ as expected and that for $\epsilon=0,$ and therefore for $\xi=0,$ it becomes independent of $p$ as stated above.

For illustrative purposes we will show a few examples of $g_p(t)$ here, labeling them explicitly as $g_{H,p}(t)$ to show the compartment size as well. When $H=2$,
i.e., when a compartment is composed of only three sites, $%
P(\epsilon )=1+\epsilon /2F$, $Q(\epsilon )=1$ and
\begin{align}
g_{2,0}(t)&=3F\left[ \vphantom{\frac{f}{F}e^{-\left( 1+\frac{f}{F}\right)
Ft}}  e^{-3Ft}-\frac{f}{F}e^{-\left( 1+\frac{f}{F}\right)
Ft}\right], \nonumber \\  \label{g2}
g_{2,\pm1}(t)&=\Delta \Bigg( \frac{3F}{2}\frac{e^{-3Ft}}{f-F} \nonumber \\&- \frac{(F^{2}-Ff/2+f^{2})e^{-(F+2f)t}}{F(f-F)}+\frac{\delta(t)}{2F} \Bigg).
\end{align}
When $H=4$ each compartment possesses 5 sites, $P(\epsilon )=1+2\epsilon
/F+\epsilon ^{2}/2F^{2}$, $Q(\epsilon )=2+\epsilon /F$ and we obtain 
\begin{align}
g_{4,0}(t)&=10F\left[ \vphantom{\frac{\sinh \left( \sqrt{5-4\frac{f}{F}+4\left( \frac{f}{F}\right) ^{2}}Ft/2\right) }{\sqrt{5-4\frac{f}{F}+4\left( \frac{f}{F}\right) ^{2}}}} e^{-\frac{5Ft}{2}}\frac{\sinh \left( \sqrt{5}Ft/2\right) 
}{\sqrt{5}} \right. \nonumber \\  &\left.-\frac{f}{F}e^{-\left( \frac{3}{2}+\frac{f}{F}\right) Ft}\frac{\sinh \left( \sqrt{5-4\frac{f}{F}+4\left( \frac{f}{F}\right) ^{2}}Ft/2\right) }{\sqrt{5-4\frac{f}{F}+4\left( \frac{f}{F}\right) ^{2}}}\right] .
\end{align}

Returning to the original $\widetilde{g}_p$ in Eq. (\ref{gepsfull}), and averaging it over all initial $p$ within the compartment with equal weight, gives a $p-$independent barrier contribution to the memory:\begin{align}
\widetilde{g}(\epsilon)&=\frac{1}{H+1}\sum_{p=-H/2}^{H/2}\widetilde{g}_p(\epsilon)=\frac{\Delta}{F\left(H+1 \right)} \nonumber \\ &\times \frac{P(\epsilon)+\left(\epsilon/2F + 2 \right)Q(\epsilon)}{P(\epsilon)+\left(\epsilon/2F+2f/F \right)Q(\epsilon)}
\label{avgdis}
\end{align}

Note that $g(t) \rightarrow 0$ as either $t\rightarrow \infty $ or $t\rightarrow 0$, since $%
\epsilon \widetilde{g}(\epsilon )\rightarrow 0$ as $\epsilon \rightarrow 0$ and $%
\epsilon \rightarrow \infty $.

\section{Continuum limit}
\label{continuum}

\begin{figure}[H]
\centering
\includegraphics[width = 8.0cm]{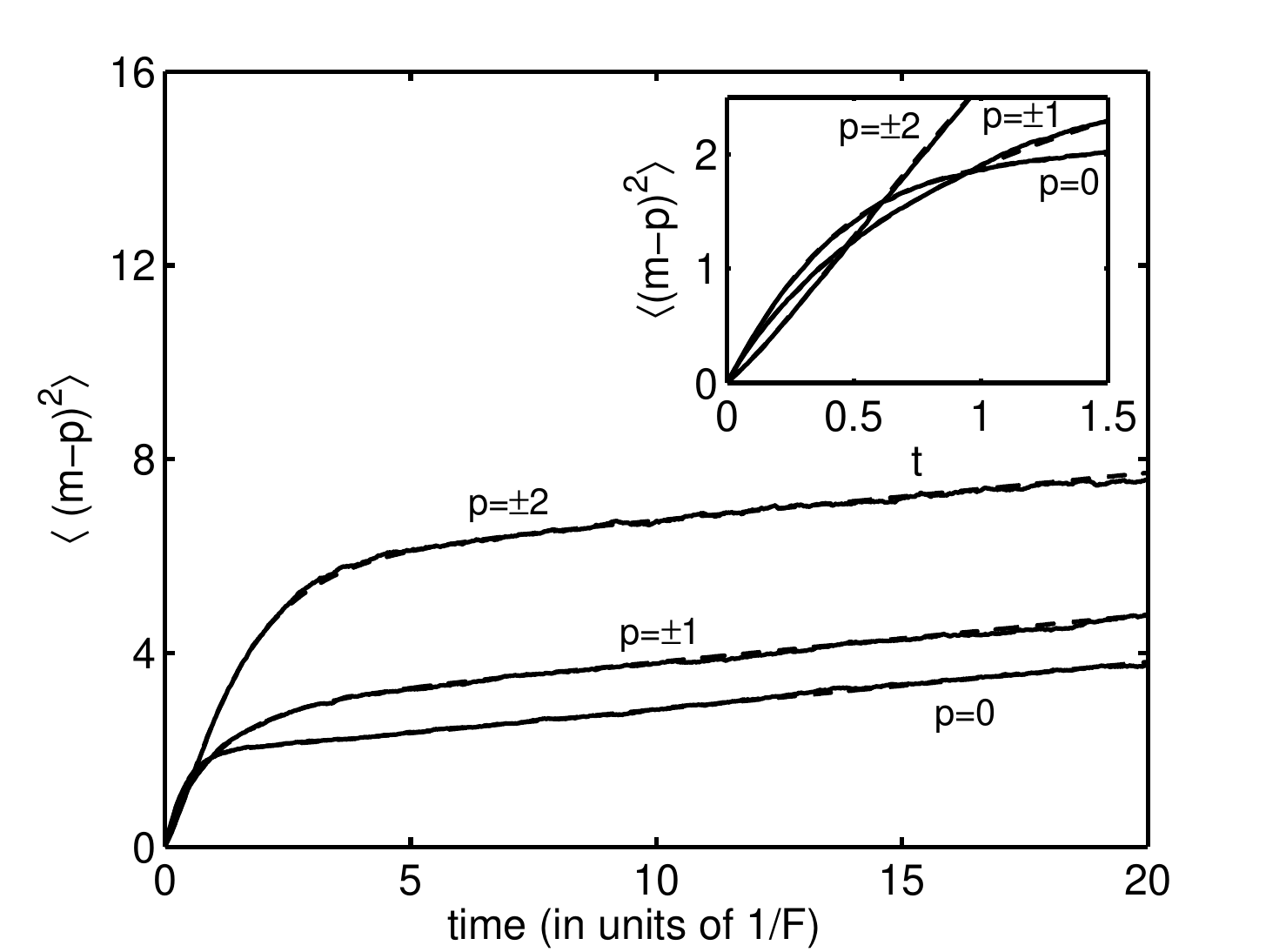}
\caption{Illustrative comparison of our analytic result for the mean square displacement with Monte-Carlo simulations, showing excellent agreement. Parameters are $f=0.01$, $F=2$ $H=4$. Solid curves show the result of the simulations, averaged over 20000 runs. Dashed lines are the analytical results. They are almost indistinguishable from one another. The inset shows the behavior at short times.}
\label{fig1}
\end{figure}

It is straightforward to compare our analytical result  with corresponding Monte-Carlo simulations of random walks with barriers. We performed such simulations by following standard procedures \cite{numrec} in a 1-D lattice with $N(H+1)$ sites  with $N$ large enough so that the molecule (random walker) never reaches the boundaries during one run of the simulation.  In Fig. \ref{fig1}, the analytical results and the results of the simulation are shown for $H=4$. We have averaged over 20000 different trajectories for each case to create the final results. Time is displayed in units of $1/F$.

While the above clear coincidence of the simulation results with our analytic solutions establishes confidence in the analysis, the comparison has been made only for  small $H$ for simplicity. Practical application of our analytic result in the cell membrane context requires an \emph{infinite} number of sites within a compartment, i.e., the continuum limit, because the molecules diffuse, rather than hop, from location to location. Therefore, we now proceed to the continuum, assuming that, as the lattice constant $a\rightarrow 0$, the products $Ha,$ $pa,$ and $fa$ tend to the respective constants $L,$ $x_0,$ and $\mathscr{D},$ but that $F$ tends to infinity an order of magnitude faster so that $Fa^2$ tends to (the diffusion constant) $D.$
Surely, $x_{0}$ is the initial position of the particle inside the central compartment and  $\mathscr{D}/D$ corresponds to the permeability
of a partially permeable barrier, indicated in ref. \cite{powlesetal} as $\mathcal{P}$. Similar continuum limits may be found in ref. \cite{simpson}. 
By retaining only the nonvanishing terms in Eq. (\ref{gepsfull}), the
continuum limit is given by
\begin{align}\widetilde{g}_{x_0}^c(s)&=\frac{s^{2}}{s\cosh s+\frac{\mathscr{D}L}{D}\sinh s}\bigg[ \coth s\cosh\left(\frac{2x_{0}}{L}s \right) \nonumber \\ &-\frac{2x_{0}}{L}\sinh\left(\frac{2x_{0}}{L}s \right) \bigg] ,
\label{ginf}
\end{align}
where the dimensionless quantity $s=\frac{L}{2}\sqrt{\frac{\epsilon}{D}}$ is proportional to the square root of the Laplace variable, the subscript $x_0$ denotes the initial location on the continuum, and the superscript $c$ is a reminder of the fact that we are treating the continuum limit. 

The continuum limit of Eq. (\ref{avgdis}) becomes (this corresponds to an average over $x_0$)
\begin{equation}
\widetilde{g}^{c}(s)=\frac{\sinh s}{s\cosh s + \frac{D_{eff}}{D-D_{eff}}\sinh s} .
\label{conavg}
\end{equation}
In terms of the inverse Laplace transform $g^{c }\left( t\right) $  of
Eq. (\ref{ginf}), the mean square displacement in the continuum limit can be written as
\begin{equation}
\left\langle (x-x_{0})^{2}\right\rangle (t)=2Dt-2D\int_{0}^{t}dt^{\prime
}\int_{0}^{t^{\prime }}g^{c}_{x_0}\left( t^{\prime \prime }\right)
dt^{\prime \prime }.  \label{msdcontinuum}
\end{equation}
Notice that $\widetilde{g}^{c }(\epsilon )$ maintains the limiting
properties of $\widetilde{g}(\epsilon )$, i.e., $\epsilon \widetilde{g}^{%
c }(\epsilon )\rightarrow 0$ as $\epsilon \rightarrow 0$ and $\epsilon \rightarrow 
\infty $ indicating that $g^{c }\left( t\right) \rightarrow 0$ as $t\rightarrow 
\infty $ and $t\rightarrow 0$. As in the discrete case, at large times it is
possible to calculate the diffusion coefficient, $$D_{eff}=D\int_{0}^{\infty %
}dt\left[ \delta (t)-g^{c }\left( t\right) \right] =D\left[ 1-%
\widetilde{g}^{c }(0)\right] ,$$ which turns out to be
\begin{equation}
D_{eff}=\frac{D}{1+\frac{D}{\mathscr{D}L}}.  \label{Deff}
\end{equation}
This result, which appears as the continuum limit of our more general discrete expressions, is equivalent to that obtained in ref. \cite{powlesetal}.

\section{Explicit Usable Expressions for the mean square displacement}
\label{expexp}

For most of the rest of the paper we will examine only the continuum limit expressions  because they allow direct comparison to experiment in the specific problem of molecular motion in cell membranes. Other systems such as those involving the hopping of excitations in a  molecular solid \cite{kenkrebook} are better treated with the help of the more general discrete case expressions we have obtained above. We will now use the notation $y_{x_0}$ to mean $\langle (x-x_{0})^{2} \rangle$ and describe how to take the inverse Laplace transform of Eq.(\ref{ginf}) to get the mean square displacement in the time domain, with the help of Eq. (\ref{msdcontinuum}). Note that the mean square displacement in the Laplace domain can be written as
\begin{gather}
\widetilde{y_{x_0}}(s)=\frac{L^{4}}{8D}\left( \frac{1}{s^{4}} - \frac{1}{s^{3}\cosh s}\zeta(s;x_{0},L) \right)  + \nonumber \\ \frac{L^{4}}{8D}\frac{D_{eff}}{D-D_{eff}}\frac{1}{s^{3}}\frac{\tanh s}{\left(\cosh s +\frac{D_{eff}}{D-D_{eff}}\sinh s  \right)}\zeta(s;x_{0},L) ,
\label{deltap}
\end{gather}
where
\begin{equation}
\zeta(s;x_{0},L)=\coth s\cosh \left( \frac{2x_{0}}{L}s\right)-\frac{2x_{0}}{L}\sinh \left( \frac{2x_{0}}{L}s\right) .
\end{equation}
Average of Eq. (\ref{deltap}) over all initial locations $x_0$ gives
\begin{align}
\widetilde{y}(s)&=\frac{1}{L}\int_{-L/2}^{L/2}dx_{0}\widetilde{\langle (x-x_{0})^{2} \rangle}(s) \nonumber \\ &=\frac{L^{4}}{8D}\left( \frac{1}{s^{4}}-\frac{\tanh s}{s^{5}} \right)\nonumber \\ &+\frac{L^{4}}{8D}\frac{1}{s^{5}}\frac{\tanh^{2}s}{\tanh s + \frac{D-D_{eff}}{D_{eff}}s}.
\label{avgsep}
\end{align}
We notice that in Eqs. (\ref{deltap}) and (\ref{avgsep}), the mean square displacement expression in the Laplace domain appears decomposed naturally as a sum of two parts. The first term represents the mean square displacement for the case of impenetrable barriers. It can be inverted easily to give a simple analytic expression in the time domain:
\begin{equation}
\frac{L^{2}}{2}\left( \frac{1}{3}-32\sum_{n=0}^{\infty}\frac{e^{-\frac{\pi^{2}}{4}(2n+1)^{2}\left(\frac{4D}{L^{2}}t \right)}}{\pi^{4}(2n+1)^{4}} \right).
\label{totally_confined}
\end{equation}
This result for the mean square displacement of a totally confined particle and related expressions for the autocorrelation function of the displacement agree with published expressions that have appeared earlier in the study of nuclear magnetic resonance microscopy and of animal motion in home ranges \cite{NMR,JTB}. 

Many problems in physics are tackled in terms of an exactly soluble system to which a perturbation is added. The choice of the partition of the system into an unperturbed and a perturbed part is obviously never unique and is itself an interesting aspect of theoretical study. This happens, for instance, in quasiparticle transport in molecular crystals and aggregates \cite{kenkrebook} wherein a localized and a delocalized part of the system represent two such possible choices. We see here that our present system displays such a choice in that one may regard the unperturbed part of our random walk problem as being either the walk on the free chain perturbed by the presence of the multiple barriers, or as the confined motion of the walker within a single well with two impenetrable barriers ($f=0$) perturbed by a leakage to the rest of the chain. In the first case, represented by Eq. (\ref{firstpartition}), the unperturbed system is represented by free motion on a chain. In the second, with the partition described by Eq. (\ref{avgsep}), it is Eq.(\ref{totally_confined}) that describes the unperturbed system. 

We now calculate the inverse Laplace transform of the full mean square displacement, focusing directly on $\widetilde{g}(\epsilon)/\epsilon^{2}$ since it is the double time integral of $g(t)$ that is required in the calculation. Here, and henceforth in this section, we drop the superscript $c$ on $g$ to avoid clutter and do not display the explicit dependence on the starting location $x_0.$  The well-known identity \cite{robertskaufmanbook}
\begin{equation}
\mathcal{L}^{-1}\left\{ \widetilde{f}(a\sqrt{\epsilon})\right\}(t)=\frac{1}{2a^{2}\sqrt{\pi \left(\frac{t}{a^{2}}\right)^{3}}}\int_{0}^{\infty}duf(u)ue^{-u^{2}\frac{a^{2}}{4t}} ,
\label{kaufid} 
\end{equation}
has found use earlier in problems similar to the present study \cite{WK}. It allows one to invert Laplace transforms of functions of the square root of the Laplace variable if the corresponding transforms of just the variable are known.

In the time domain, we have
\begin{align}
y (t)&=2Dt-2D\int_{0}^{t}dt^{\prime
}\int_{0}^{t^{\prime }}dt^{\prime \prime } \nonumber \\ &\times \oint_{C}d\epsilon \widetilde{g}\left( \epsilon \right)e^{\epsilon t^{\prime \prime }} ,
\label{MSDT}
\end{align}
where $C$ is the Bromwich contour enclosing all the singularities of $\widetilde{g}\left( \epsilon \right)$ (note that all of the singular points of Eq. (\ref{ginf}) are on the imaginary axis). After performing some simple (although tedious) algebraic manipulations, employing the notation $\gamma$=$D_{eff}~/~(~D~-~D_{eff}~)$, $\alpha$=$2x_{0}/L$, $\tau$=$4Dt/L^{2}$, and using Eq. (\ref{kaufid}), we get
\begin{equation}
y(\tau)=\frac{L^{2}}{2}\left( \tau - \left( \sigma_{1}(\tau) - \gamma\sigma_{2}(\tau) - \alpha\sigma_{3}(\tau) \right) \right) ,
\end{equation}
where
\begin{gather}
\sigma_{1}(\tau)=\frac{1}{2\pi i}\oint_{C} ds \frac{\cosh(\alpha s)e^{s^{2}\tau}}{s^{2}\sinh s}, \nonumber \\
\sigma_{2}(\tau)=\frac{1}{2\pi i}\oint_{C} ds \frac{\cosh(\alpha s)e^{s^{2}\tau}}{s^{2}\left( s\cosh s+\gamma\sinh s \right)}, \nonumber \\
\sigma_{3}(\tau)=\frac{1}{2\pi i}\oint_{C} ds \frac{\sinh(\alpha s)e^{s^{2}\tau}}{s\left( s\cosh s+\gamma\sinh s \right)}.
\label{sig3}
\end{gather}
Therefore $\sigma_{1}$, $\sigma_{2}$, $\sigma_{3}$ will be equal to the sum of the residues of the integrands in Eqs. (\ref{sig3}). Let 
\begin{gather}
w_{1}(s,\tau)=\frac{\cosh(\alpha s)e^{s^2\tau}}{s^2\sinh s}, \nonumber \\
w_{2}(s,\tau)=\frac{\cosh(\alpha s)e^{s^2\tau}}{s^{2}(s\cosh s+\gamma \sinh s)}, \nonumber \\
w_{3}(s,\tau)=\frac{\sinh(\alpha s)e^{s^2\tau}}{s(\cosh s+\gamma\sinh s).} \nonumber \\
\end{gather}
Then,
\begin{align}
y(\tau)&=\frac{L^{2}}{2}\bigg( \tau -\bigg(\sum_{s_n} Res\{w_{1},s_{n}\} \nonumber \\
&- \gamma \sum_{s_n} Res\{w_{2},s_{n} \} - \alpha \sum_{s_n} Res\{ w_{3},s_{n}\} \bigg) \bigg),
\end{align}
where $s_n$'s are the solutions of $-s=\gamma\tanh s.$ At $s=0$, $w_{1}$ and $w_{2}$ have poles of order 3 and $w_{3}$ has a pole of order 2. The corresponding residues are $\tau +\alpha^{2}/2 - 1/6$, $(6\tau(\gamma+1)+3\alpha^{2}(\gamma+1)-(\gamma+3))/(6(\gamma+1)^{2})$ and $\alpha/(\gamma+1)$ for $w_{1}$, $w_{2}$ and $w_{3}$ respectively. Note that each of the functions $w_{1}$, $w_{2}$ and $w_{3}$ has infinitely many poles besides the pole at $s$=0. The rest of the poles of $w_{1}$ are all simple poles (as can easily be seen by Taylor-expanding $\sinh s$) located at $s$=$im\pi$ where $m=\pm1, \pm2, \pm3...$ . For $w_{2}$ and $w_{3}$, we cannot find the exact locations of the poles analytically because of the need to solve a transcendental equation. Although we do not have an analytical expression for the location of these poles, we know that they are all simple poles because the first derivative of $s\cosh s+\gamma \sinh s$ does not vanish at  $s_n$. Therefore, excluding the pole at $s_{n}=0$ in the $n$-summations below,
\begin{align}
\sigma_{1}(\tau)&=\tau +\alpha^{2}/2 - 1/6+\sum_{s_n} \lim_{s\rightarrow s_{n}}\frac{\cosh(\alpha s)e^{s^{2}\tau}}{s^{2}\cosh s}, \label{lsig1} \\
\sigma_{2}(\tau)&=\frac{6\tau(\gamma+1)+3\alpha^{2}(\gamma+1)-(\gamma+3)}{6(\gamma+1)^{2}} \nonumber \\ &+\sum_{s_n} \lim_{s\rightarrow s_{n}}\frac{\cosh(\alpha s)e^{s^{2}\tau}}{s^{2}(s\sinh s +(1+\gamma)\cosh s)}, \label{lsig2} \\
\sigma_{3}(\tau)&=\frac{\alpha}{(\gamma+1)}+\sum_{s_n} \lim_{s\rightarrow s_{n}}\frac{\sinh(\alpha s)e^{s^{2}\tau}}{s(s\sinh s +(1+\gamma)\cosh s) }. 
\label{lsig3}
\end{align}
The roots of $1/\sinh(s)$ can be found analytically, and $\sigma_{1}(\tau)$ can be expressed as 
\begin{equation}
\sigma_{1}(\tau)=\sum_{m=-\infty, m\neq 0}^{\infty}(-1)^{m+1}\frac{\cos(\alpha m \pi)e^{-m^{2}\pi^{2}\tau}}{m^{2}\pi^{2}},
\end{equation}
the right hand side being simply related to the $\tau$-integral of elliptic theta functions.

Because we cannot solve $-s=\gamma \tanh s$ exactly, it is not possible to write analytical expressions for $\sigma_{2}(\tau)$ and $\sigma_{3}(\tau)$. However, we can find the roots $s_{n}$ of that equation numerically with high precision, and evaluate the sums in Eqs. (\ref{lsig2}) and (\ref{lsig3}) to get $\sigma_{2}(\tau)$ and $\sigma_{3}(\tau)$. In doing this, we use the bisection method to find the first few thousands of the roots of $\sigma_{2}(\tau)$ with an accuracy of $10^{-13}$. Because of the factor $e^{s^2\tau}$ in Eqs. (\ref{lsig2}) and (\ref{lsig3}), the sums converge very fast as $s_{n}$'s are purely imaginary. Note that $\lim_{s\rightarrow s_{n}}\cosh(\alpha s)$ and $\lim_{s\rightarrow s_{n}}\sinh(\alpha s)$ lie in $[-1,1]$ and the magnitude of $\lim_{s\rightarrow s_{n}}((1+\gamma)\cosh s +s \sinh s)$ tends to $\infty$ with increasing Im$(s_{n})$. As a result, it is possible to obtain accurate results without summing over a large number of residues.  A similar procedure can be followed \footnote{We can gladly provide a MATLAB code, that performs the tasks described above in a few seconds, to anyone interested.} to find the inverse Laplace transform of Eq. (\ref{avgsep}).  

\begin{figure}
\centering
\includegraphics[width = 8.0cm]{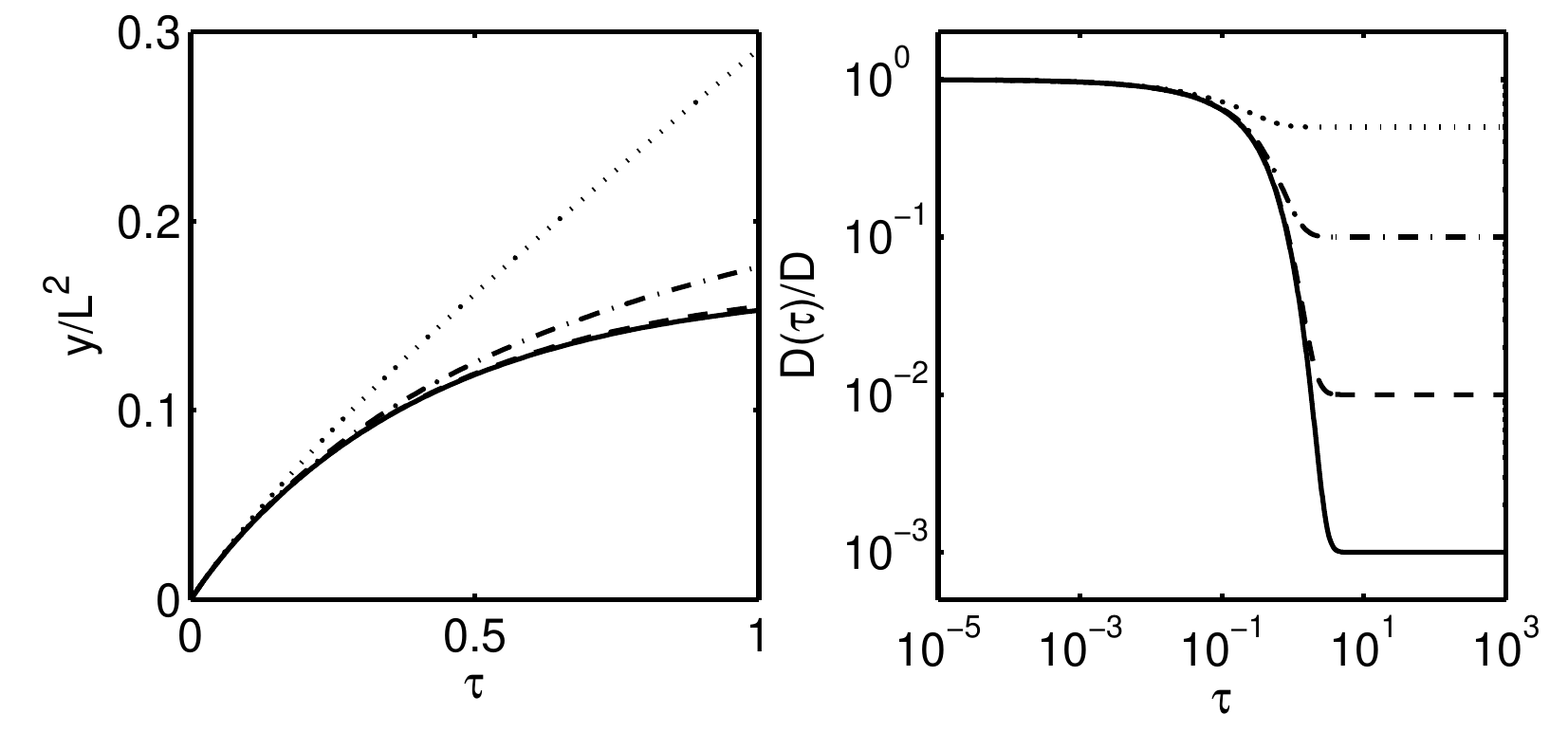}
\caption{Time dependence of the mean square displacement and the instantaneous diffusion coefficient for various magnitudes of the confinement effect. Plotted as a function of time is the normalized y(t)  (left) and $D(t)/D$   (right), averaged over all initial locations in the compartment, for different values of $D_{eff}/D$: 0.001 (solid line), 0.01 (dashed line), 0.1 (dash-dotted line), 0.5 (dotted line). All quantities considered are made dimensionless appropriately.}
\label{avg}
\end{figure}

We have displayed several of the details of the inversion procedure we have used because of its involved nature and because it is not often that (even partial) inversions resulting in explicitly usable time domain expressions are possible in similar problems. In Fig. \ref{avg} we show the dependence, on the (dimensionless) time $\tau$, of the normalized instantaneous diffusion coefficient, 
\begin{equation}
\frac{D(t)}{D}=\frac{1}{2D}\frac{d}{dt}y(t)=\frac{2}{L^{2}}\frac{d}{d\tau}y(\tau). 
\label{dtd}
\end{equation}

\section{comparison with experiment}
\label{experiment}
In comparing our theoretical results with experimental observations in the cell membrane field, we used two sets of data by digitizing two time plots of the molecular mean square displacement available in the published literature. The first is Fig 4b (left) in ref. \cite{suzukietal} which is about the diffusion of a gold-tagged G-protein coupled $\mu$-opioid receptor; the second is Fig 2b in ref. \cite{fujiwaraetal} which is about the diffusion of a phospholipid molecule (1,2-dioleoylsn-glycero-3-phosphoethanolamine). Both sets of observations refer to normal rat kidney cells. The observations were made at a time resolution of $25 \mu s$ which is high enough to capture the change in the diffusion constant at times long and short compared to expected values of  $L^{2}/4D$. 

We first fitted our $y$ vs $t$ curve, given by Eq. (\ref{MSDT}), to the data in Fig 4b of ref. \cite{suzukietal}, both to the x and y components of the mean square displacement, and extracted the compartment sizes  through standard procedures. It is not possible to make accurate deductions because the data in the literature are available only for single trajectories rather than for ensemble averages for a large number of them. Within this limitation, we have been able to conclude by the application of our theory to the data that the linear extent of the confining compartment should lie between $250 nm$ and $470 nm$. Our conclusion is in agreement with the distribution of compartment sizes given in Fig 4d of ref. \cite{suzukietal} and generally with values discussed in ref. \cite{kusumireview}.

\begin{figure}
\centering
\includegraphics[width = 8.0cm]{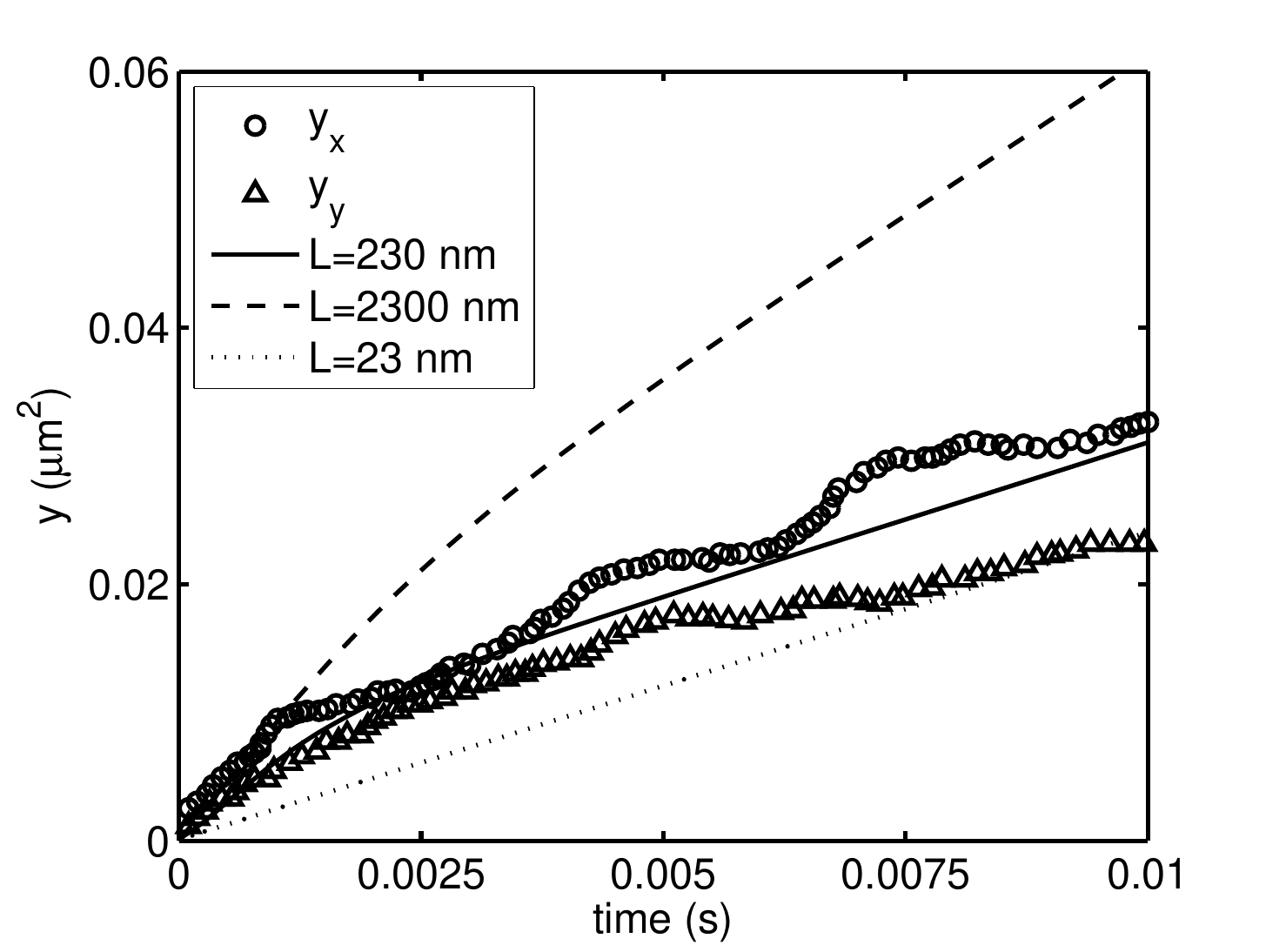}
\caption{Comparison of diffusion observations of phospholipids reported in ref.  \cite{fujiwaraetal} to our theoretical predictions based on 3 disparate values of the compartment size $L$ used in our  Eq. (\ref{MSDT}). One of the values used is that deduced in ref. \cite{fujiwaraetal}, $230\ nm$. There is very good agreement of theory (solid line) and observations as shown. Each of the other two values differs by an order of magnitude: $2300\ nm$ (dashed line) and $23\ nm$ (dotted line). In both these cases theory shows poor agreement with experiment.  We have used $D$=$4.6\ \mu m^{2}s^{-1}$ and $D_{eff}$=$1.2\ \mu m^{2}s^{-1}$.}
\label{differentL}
\end{figure}

A different system we examined is that involving the diffusion of smaller transmembrane molecules than proteins: phospholipids \cite{fujiwaraetal}. By evaluating Eq. (\ref{MSDT}) using experimentally deduced values for $D$, $D_{eff}$ and $L$, we verified that conclusions derived in ref. \cite{fujiwaraetal} are in agreement with theory.  This is displayed in Fig. \ref{differentL}. To show the agreement clearly, we use for the compartment size, $L$, a substantially different value from that deduced in ref. \cite{fujiwaraetal}, $230\ nm$, and point out that the $y$ vs $t$ curves found by using Eq. (\ref{MSDT}) do not then agree with the experiment at all. 

\begin{figure}
\centering
\includegraphics[width = 8.0cm]{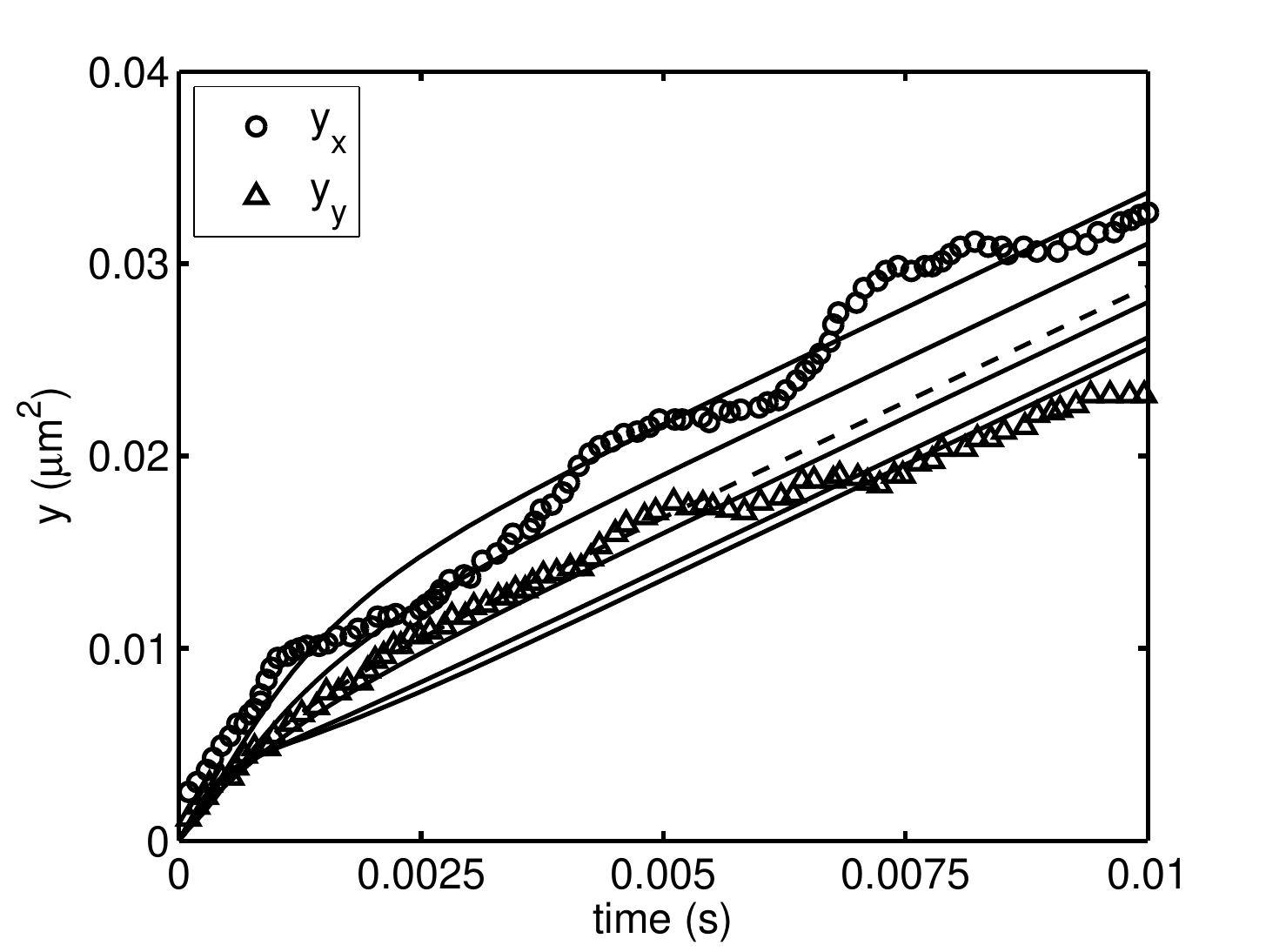}
\caption{Theoretical predictions for various initial placements of the molecule compared to observations reported in ref. \cite{fujiwaraetal}. Circles and triangles denote observations. Lines represent our theory. Dashed line shows the average of the mean square displacement over all initial positions. For the solid lines, the initial position is: $x_{0}$=$0$, $L/8$, $L/4$, $3L/8$ and $7L/16$, going upwards from the lowest curve in the plot. Values of $D$ and  $D_{eff}$ that correspond to the observations are as in Fig. \ref{differentL} and the deduced compartment size is $L$=$230\ nm$. }
\label{init}
\end{figure}

As the data corresponds to a single trajectory, it is worthwhile to explore the effects of the initial position of the molecule on the dynamics. In Fig. \ref{init}, we show the effect of the initial position of the particle on the $y$ vs $t$ curve. The parameters $D$, $D_{eff}$ and $L$ are the same as the ones reported in ref. \cite{fujiwaraetal}. For illustrative purposes we also show the $y$ vs $t$ curve averaged over all initial positions as the dashed line in Fig. \ref{init}.

\section{Concluding Remarks}
\label{conclusion}

The investigation reported in this paper has had two separate aims. One is to develop an efficient and useful formalism to describe the motion of a transmembrane molecule on a cellular membrane following ideas inherent in the current  membrane skeleton fence model of Kusumi and collaborators \cite{kusumireview}. The other aim is to understand, in general, the problem of a random walker moving in a region with confinement. Working in a 1-D system for simplicity, we gave the exact solution of the probabilities in the Laplace domain for a discrete chain, derived from them analytic solutions for the mean square displacement of the molecule, and introduced memory functions to understand the evolution physically. We achieved a detailed understanding of the motion arising from repeated encounters of the random walker with the barriers and dependence on initial placement of the walker. These discrete solutions are of use not only to the present problem but also to diverse contexts such as the motion of excitations in crystals and molecular aggregates. 

We took the continuum limit of our results to obtain them in terms of parameters that can be measured experimentally in the context of the cellular membrane and presented a procedure to get explicit and easily usable expressions in the time domain. While a piece of the Laplace inversion  has to be done either approximately or numerically, other physically relevant parts are implemented fully analytically, the entire calculation being quite usable from a practical standpoint. Our expressions are more useful in this regard than those available earlier, e.g., in ref. \cite{powlesetal}. Specifically, our results may be typified by the prescription for the instantaneous diffusion coefficient normalized to the diffusion constant, i.e., $D(t)/D$ given in Eq. (\ref{dtd}). Explicitly, identifying the $m$-summation \cite{gradhik} $$\sum_{m=-\infty}^{\infty} (-1)^{m} \cos(\alpha \pi m) e^{-\pi^{2} \tau m^{2}} $$ with  the elliptic theta function of the fourth kind, $\varthetaup_{4}$, as given in Eq. (8.180) of ref. \cite{gradhik}, we can write,

\begin{gather}
\frac{D(\tau)}{D} = \frac{D_{eff}}{D}+1 -\varthetaup_{4} \left(\alpha \pi /2 \mid i\pi\tau \right) \nonumber \\ 
+ \sum_{s_n,s_n \neq 0}\lim_{s \rightarrow s_n}   e^{s^{2} \tau} \frac{ \cosh(\alpha s)}{\cosh s} \left[ \frac{\frac{D_{eff}}{D-D_{eff}}+\alpha s \tanh (\alpha s)}{\frac{D}{D-D_{eff}}+s \tanh s}\right]. 
\label{final}
\end{gather}
Here  $\alpha$, defined earlier as $2x_0/L$, measures the initial location relative to the compartment size, and $\tau=4Dt/L^2$ is a dimensionless time. At infinite time, $\tau \rightarrow \infty$, since the theta function tends to $1$ and the summation over the $s_n$'s vanishes because of the exponential factor (note that that the $s$'s lie on the imaginary axis), $D(\tau)$ becomes the effective diffusive constant $D_{eff}$. At the initial time, on the other hand, the theta function vanishes and $D(\tau)$ takes the value $D$ for all initial placements of the molecule not at the edge, $x_0 \neq L/2$. The mean square displacement is obtained trivially from the integral of the right hand side of Eq. (\ref{final}). Of the three terms to be computed in Eq. (\ref{final}), the first two are given analytically in our theory and the last is obtained numerically.

Our expressions are characterized by three parameters: $L$, the compartment size, $D$,  the initial diffusion constant (as $t\rightarrow 0$), and $D_{eff}$, the effective diffusion constant (as $t\rightarrow \infty$). We showed how our theory can be used directly to deduce the compartment size $L$ by fitting the analytic expressions  to data obtained in single particle tracking experiments \cite{kusumireview, suzukietal}. 

\begin{figure}
	\begin{center}
	\includegraphics[width=8cm]{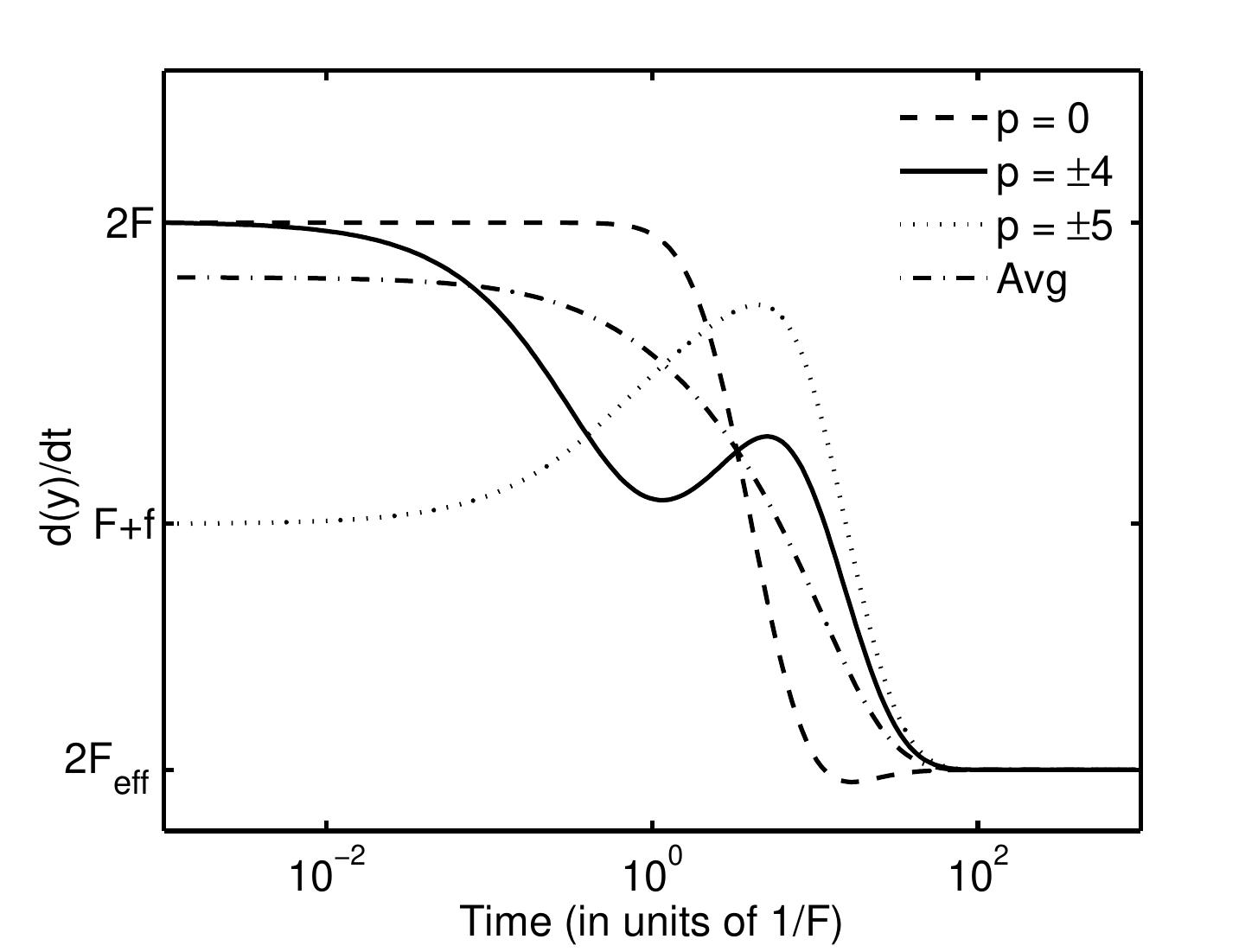}
	\caption{Effect of initial placement of the random walker at site $p$ within a compartment on the time dependence of the effective transfer rate which, in the continuum limit, would be proportional to the time-dependent effective diffusion coefficient. Each compartment has 11 sites, i.e., $H=10$, and  $f/F=0.01$. Averaging equally over all initially localized placements results, as shown by the dash-dotted line, in a monotonically decreasing transfer rate. However, interesting structures appear for initial placements at the center of, end of, and elsewhere in the compartment as shown. See text for explanation.}
		\label{doft}
		\end{center}
\end{figure}

An interesting outcome of our analysis is the description of the explicit influence of the initial placement of the random walker. The influence is observed in the time dependence of the mean square displacement. The dependence is evidently a consequence of the relative location of the barriers and the site of initial placement. With the help of the analytical expressions we have obtained, we plot in Fig. \ref{doft}, the instantaneous time derivative of the mean square displacement, a quantity that is proportional to the instantaneous diffusion coefficient $D(t)$ in the continuum limit, for several cases of the initial placement $p$ of the molecule within the compartment. We show the discrete case so that the results can be carried over to other physical contexts such as excitation transfer as well. We observe that the time derivative of the mean square displacement does not simply change monotonically from an initial value to a lower final value. Instead, interesting structures appear.

The compartment considered in the plot has a total of $11$ sites. For central initial placement ($p=0$), we see (this is the top curve shown dashed) the appearance of a curious dip at the bottom of the curve. The dip means that for some time the molecule diffuses slower than what it does eventually via the final effective diffusion constant. The situation changes quite a bit as we vary the site of initial placement, although the final effective diffusion constant always remains unaffected. For initial placement at the edge of the compartment, denoted in Fig. \ref{doft} by the dotted curve ($p=\pm5$), the molecule begins with an effective transfer rate which is not $2F$ as in other cases but a lower value $F+f.$ It then \emph{rises} first, reaches a peak, and then decreases to the final effective value. For initial placement which is neither at the center of the compartment nor at its edge, shown in the plot by the solid curve ($p=\pm4$), we see that the effective motion occurs with an initial transfer rate $2F$ and decreases, both features being shared with the case of central placement, but that there is a subsequent increase to a peak and decrease to the final effective value, features shared with the case of edge placement. 

It is easy to understand all these features. The molecule, if placed centrally within the compartment, tends to move initially with the transfer rate or diffusion constant characteristic of the barrier-less system until it meets the barrier. At this point it crosses the barrier more slowly and the effective transfer rate drops. When the molecule has diffused to the next compartment it is outside the immediate influence of  the barrier and the effective diffusion is therefore faster. The combined effect of repeated free diffusion and barrier-hindered diffusion eventually brings the effective diffusion constant to its long time value. If, however, the initial placement of the molecule is at the compartment edge, it already begins moving with an effective diffusion constant lower than the free value because of the immediate effect of the barrier. It diffuses faster from then on until other barrier encounters including the one at the other edge of the initial compartment decrease the rate of diffusion. For intermediate initial placements these effects happen one after the other as the molecule encounters first the barrier on one side and then the one on the other side of the compartment. 

Whether the structures described are visible in a specific kind of experiment will obviously depend on the freedom one has in initial placement of the random walker. We believe in light of the data that we have seen in the cell membrane field that the average alone matters and that the average, as shown in the plot, shows merely a monotonic decrease of the effective diffusion constant. The dips have been seen in the theoretical work of Powles et al \cite{powlesetal}. However, no explanation has been given. Indeed, (see Fig. 3 of ref. \cite{powlesetal} and the subsequent discussion) the authors of that work referred to the dip as an ``unexpected minimum'' and did not connect it to the dynamics we have described above. 

Missing from our present analysis are the effects of disorder in the placement and height of the barriers, i.e., in the \emph{random} nature of $H$ and $f$ in real systems. Static disorder effects of this nature are being analyzed by us in several different ways, including the replacement of the periodic expressions for the quantity $\mu$ used in the present paper by random counterparts that take into account probability distributions, and the use of effective medium theories. These analyses will be reported in a separate publication. Of interest to us are also
calculations of random walks in the presence of \emph{dynamic} fences that allow us to address properly the case where the moving molecules are the lipids while the fences are made by the proteins (as has been suggested in the literature \cite{kusumireview}, and calculations which are valid in the presence of intermolecular interactions.

\begin{acknowledgements}
This work was supported in part by the NSF under
grant no. INT-0336343, by NSF/NIH Ecology of In-
fectious Diseases under grant no. EF-0326757, by the
Program in Interdisciplinary Biological and Biomedical
Sciences at UNM funded by the Howard Hughes Medi-
cal Institute, and by DARPA under grant no. DARPA-
N00014-03-1-0900. We acknowledge helpful discussions with Paul E. Parris.
\end{acknowledgements}

\appendix

\section{}
\subsection{Evaluation of $\mu$}
Exploiting the linearity of Eq. (\ref{starting}), it is possible to write
its solution in the Laplace domain as 
\[
\widetilde{P}_{m}(\epsilon )=\widetilde{\eta }_{m}(\epsilon )-\Delta \sum_{r}^{\prime}%
\widetilde{p}_{r}(\epsilon )\left[ \widetilde{\Psi }_{m-r}(\epsilon )-%
\widetilde{\Psi }_{m-r-1}(\epsilon )\right] , 
\]
where we denote the difference between the probability to the left and right
of the barriers $P_{r+1}(t)-P_{r}(t)$ by $p_{r}(t)$ and its Laplace
transform by $\widetilde{p}_{r}(\epsilon )$. To find an explicit solution
from the above equation, we first write its particular cases for $m=s$ and $%
m=s+1$ where $s$ and $s+1$ mark, respectively, the sites to the left and to
the right of a barrier: 
\begin{eqnarray}
\widetilde{P}_{s+1}(\epsilon ) =\widetilde{\eta }_{s+1}(\epsilon )-\Delta
\sum_{r}^{\prime}\widetilde{p}_{r}(\epsilon )\left[ \widetilde{\Psi }
_{s-r+1}(\epsilon )-\widetilde{\Psi }_{s-r}(\epsilon )\right] ,  \nonumber \\
\widetilde{P}_{s}(\epsilon ) =\widetilde{\eta }_{s}(\epsilon )-\Delta
\sum_{r}^{\prime}\widetilde{p}_{r}(\epsilon )\left[ \widetilde{\Psi }_{s-r}(\epsilon
)-\widetilde{\Psi }_{s-r-1}(\epsilon )\right] \nonumber \\ \label{pspsplusone}
\end{eqnarray}
and then subtract the second expression in Eq. (\ref{pspsplusone}) from the
first one to get the difference 
\begin{equation}
\widetilde{p}_{s}(\epsilon )=\widetilde{\zeta }_{s}(\epsilon )-\Delta
\sum_{r}^{\prime}\widetilde{p}_{r}(\epsilon )\widetilde{\varphi }_{s-r}(\epsilon ),
\label{forFT}
\end{equation}
where $\widetilde{\zeta }_{s}(\epsilon )=\widetilde{\eta }_{s+1}(\epsilon )-%
\widetilde{\eta }_{s}(\epsilon )$, and 
\begin{equation}
\widetilde{\varphi }_{r}(\epsilon )=\widetilde{\Psi }_{r+1}(\epsilon )+%
\widetilde{\Psi }_{r-1}(\epsilon )-2\widetilde{\Psi }_{r}(\epsilon )=\frac{1%
}{F}\left[ \epsilon \widetilde{\Psi }_{r}(\epsilon )-\delta _{r,0}\right] .
\label{phi}
\end{equation}
In the last step we have used the original equation of motion to simplify
the propagator expression.

If (\ref{forFT}) can be solved for $\widetilde{p}_{s}(\epsilon )$ for all
sites associated with the barrier placements, an explicit solution for the
probabilities of \emph{all} sites $m$ can be obtained. This is the general idea of this development. Such a solution can be obtained exactly for one or a few barriers
through an explicit evaluation of determinants of manageable size involved
in the simultaneous equations (\ref{forFT}). In the case of random and
periodic arrangement of barrier positions $\widetilde{p}_{s}(\epsilon )$, can
also be calculated explicitly as follows.

Summation of (\ref{forFT}) over barrier locations $s$ gives, exactly, 
\[
\sum_{s}\widetilde{p}_{s}(\epsilon )=\sum_{s}\widetilde{\zeta }_{s}(\epsilon
)-\Delta \sum_{s}\widetilde{\mu }_{s}(\epsilon )\widetilde{p}_{s}(\epsilon
), 
\]
where the new function $\widetilde{\mu }_{s}(\epsilon )$ is defined through 
\[
\widetilde{\mu }_{s}(\epsilon )=\sum_{r}^{\prime}\widetilde{\varphi }_{s-r}(\epsilon
). 
\]
This $\mu $-function is the sum of propagators among sites placed at the left end of the periodically placed barriers. If the barriers are placed at random
positions, the average can be assumed to be independent of the location $s$,
and $\widetilde{\mu }_{s}(\epsilon )=\widetilde{\mu }(\epsilon )$. An ensemble average is envisaged here. Then,
\[
\sum_{s}\widetilde{p}_{s}(\epsilon )=\frac{\sum_{s}\widetilde{\zeta }
_{s}(\epsilon )}{1+\Delta \widetilde{\mu }(\epsilon )}, 
\]
is an explicit solution.

For periodic arrangements of barriers, as used in earlier treatments of the problem \cite{powlesetal}, the wall location summation given by 
\begin{equation}
\widetilde{\mu }_{s}=\sum_{r}^{\prime}\widetilde{\varphi }_{s-r}=\frac{1}{F}
\sum_{r}^{\prime}\left( \epsilon \widetilde{\Psi }_{s-r}-\delta _{s-r,0}\right) ,
\label{mus}
\end{equation}
can now be done and the result shown to be independent of $s.$ As stated in
the introduction, the sites to the left of each barrier are located at $%
r=H/2+(H+1)l$ with $l$ varying as an integer from $-\infty $ to $+\infty $.
We can thus change the summation to $s-r=(H+1)l$ and calculate

\begin{equation}
\widetilde{\mu }=\frac{1}{F}\sum_{l=-\infty }^{+\infty }\left( \epsilon 
\widetilde{\Psi }_{l}-\delta _{l,0}\right) .  \label{musum}
\end{equation}
To proceed with the evaluation of the sum we use the fact that the propagators for
chain involve modified Bessel functions. It follows that 
the sum in Eq. (\ref{mus}) is a geometric sum with 
\begin{eqnarray}
\widetilde{\mu }&=&\frac{1}{F}\left( \frac{\epsilon \sum_{l=-\infty }^{+\infty
}e^{-\xi (H+1)\left| l\right| }}{2F\sinh \xi }-1\right) \nonumber \\ &=&\frac{1}{F}\left[ 
\frac{\tanh \left( \xi /2\right) }{\tanh \left( \xi \left( H+1\right)
/2\right) }-1\right] .  \label{mu}
\end{eqnarray}

\subsection{Evaluation of Two Sums}
\label{2sums}

In order to evaluate Eq. (\ref{g}), we need to perform the following two sums
\begin{align}
&\sum_{r}^{\prime}\left( \widetilde{\Psi }_{r-p}-\widetilde{\Psi }
_{r+1-p}\right) , \label{sum1} \\
&\sum_{r}^{\prime}r\left( \widetilde{\Psi }_{r-p}-\widetilde{\Psi }
_{r+1-p}\right) . \label{sum2}
\end{align}
The first can be shown, after some algebra, to be\begin{equation}
\sum_{r}^{\prime}\left( \widetilde{\Psi }_{r-p}-\widetilde{\Psi }
_{r+1-p}\right)=\frac{\sinh(\xi/2)}{F\sinh(\xi)}\frac{\sinh(\xi p)}{\sinh(\xi(H+1)/2)} ,
\label{sum1f}
\end{equation}
whereas the second yields
\begin{gather}
\sum_{r}^{\prime}r\left( \widetilde{\Psi }_{r-p}-\widetilde{\Psi }
_{r+1-p}\right)=\nonumber \\ \frac{H+1}{2F\sinh\xi}\frac{\sinh(\xi/2)\cosh\left(\frac{\xi(H+1-2p)}{2} \right)}{\sinh^{2}\left(\frac{\xi (H+1)}{2}  \right)}.
\label{sum2f}
\end{gather}
Now we can use Eqs. (\ref{sum1f}) and (\ref{sum2f}) in Eq. (\ref{g}) to get
\begin{align}
\widetilde{g}_{H,p}(\epsilon)&=\frac{\Delta}{F}\frac{\epsilon}{1+\frac{\Delta}{F}\left( \frac{\tanh(\xi/2)}{\tanh(\xi(H+1)/2)}-1 \right)} \nonumber \\
&\times\Bigg[\frac{(H+1)\sinh(\xi/2)}{2F\sinh(\xi) \sinh(\xi(H+1)/2)} \nonumber \\ &\times\left( \frac{\cosh(\xi(H+1)/2)\cosh(\xi p)}{\sinh(\xi(H+1)/2)}-\sinh(\xi p) \right) \nonumber \\
&+\frac{H+1-2p}{2}\left( \frac{\sinh(\xi/2)\sinh(\xi p)}{F\sinh(\xi)\sinh(\xi(H+1)/2)} \right) \Bigg] .
\end{align}
Finally, using the expansion of the hyperbolic cotangent leads to Eq. (\ref{gepsfull}).

\end{document}